\documentstyle[12pt]{article}
\newcommand{\quattrova}{($\varphi^{a},c^{a},\lambda_{a},{\bar c}_{a}$)}

\begin{document}

\baselineskip =15.5pt
\pagestyle{plain}
\setcounter{page}{1}

\begin{titlepage}

\begin{flushright}
\end{flushright}
\vfil

\begin{center}
{\huge A New Look at the\\ 
Schouten-Nijenhuis, Fr\"olicher-Nijenhuis\\
and Nijenhuis-Richardson Brackets for Symplectic Spaces}
\end{center}

\vfil

\begin{center}
{\large E. Gozzi$^{\dag}$ and D. Mauro}\\
\vspace {1mm}
Dipartimento di Fisica Teorica, Universit\`a di Trieste, \\
Strada Costiera 11, P.O.Box 586, Trieste, Italy \\ $^{\dag}$ and INFN, Sezione 
di Trieste.\\
\vspace {1mm}
\vspace{3mm}
\end{center}

\vfil

\begin{center}
{\large Abstract}
\end{center}

\noindent
In this paper we re-express the Schouten-Nijenhuis, the Fr\"olicher-Nijenhuis
and the Nijenhuis-Richardson brackets on a symplectic space
using the extended Poisson\break brackets structure present in the path-integral

formulation of classical mechanics.
\vfil
\end{titlepage}
\newpage
\section{Introduction}

Some time ago a {\it path-integral} formulation of classical 
mechanics (CM) appeared in the literature \cite{Gozzi}.  This formulation
was nothing else than the path-integral counterpart of the {\it operatorial}
version of CM provided long ago by Koopman and von Neumann~\cite{Koop}.
From now on we will refer to the formulation contained in ref.~\cite{Gozzi} 
as  CPI for "{\bf C}lassical {\bf P}ath  {\bf I}ntegral".

Calling the phase-space of the system as ${\cal M}$  one had
that in the CPI, besides the 2n phase-space variables:
$\varphi^{a}=(q^{1}\cdots q^{n},p^{1}\cdots
p^{n})$, the measure in the path-integral contained a set of 6n auxiliary 
variables which were indicated as $(\lambda_{a},c^{a},{\bar c}_{a})$. All
together these 8n variables $(\varphi^{a},\lambda_{a}, c^{a},{\bar c}_{a})$
labelled a space whose geometrical meaning was studied in ref.~\cite{Regini}.
It turned to be what is called  the cotangent bundle to the reversed-parity 
tangent bundle to
phase-space\footnote{For the definitions and meanings of
these words we refer the reader to ref. \cite{Regini}.}, and is indicated 
in brief as $T^{\star}(\Pi T{\cal M})$. 
Being a cotangent bundle this space had a Poisson structure which was
called~\cite{Gozzi} {\bf E}xtended {\bf P}oisson {\bf B}rackets (or {\it EPB})
to distinguish it from the standard Poisson brackets defined on the 
phase-space ${\cal M}$. Via these {\it EPB} and the 8n variables indicated above
it was shown~\cite{Gozzi} that all the standard variables (forms, multivectors,
 etc) could be mapped into functions of our
8n variables and the standard  operations (exterior derivative, 
interior contraction, Lie-brackets, Lie-derivative, etc.)
of the Cartan calculus~\cite{Marsd} could be obtained by inserting those
functions into chains of {\it EPB}.

What  had not been mapped into this formalism of the CPI and of the {\it EPB}
were those generalizations
of the Lie-brackets known~\cite{Kolar}\cite{vaisman}  as the Schouten-Nijenhuis ({\bf SN}) 
brackets,  the Fr\"olicher-Nijenhuis ({\bf FN}) brackets and the 
Richardson-Nijenhuis ({\bf RN}) ones. In this paper we will derive the 
above mentioned mapping for these backets.

The paper is organized as follows. In Sec. {\bf 2} we will briefly review the 
path integral  for classical mechanics (CPI) and explain the {\it EPB} 
structure present there.
In Sec. {\bf 3} we will show how to do the Cartan calculus via our variables
and the associated {\it EPB} structure.
In Sec. {\bf 4} we will map the {\bf SN} brackets, the {\bf FN} brackets and the 
{\bf RN} ones into operations done with only the {\it EPB} brackets
with inserted different functions of our variables.
In Appendix A we will report the main formulas
of the Cartan calculus while the calculations of Sec. 4 are 
given in details in Appendices B, C and D.
\section{Path Integral for Classical Mechanics (CPI)}

The idea is to give a {\it path integral} for CM  which will reproduce the 
{\it operatorial} version of CM  as given
by the {\it Liouville}  operator~\cite{Koop} or by the {\it Lie derivative} 
of the Hamiltonian flow~\cite{Marsd}.
We will be brief here because more details can be found in \cite{Gozzi}.

Let us start with a 2n-dimensional phase space ${\cal M}$ whose coordinates
are indicated as $\varphi^{a}$$(a=1,\ldots,2n)$, i.e.: $\varphi^{a}=(q^1\cdots q^n,p^1\cdots,p^n)$. Let us 
indicate the 
Hamiltonian of the system as $H(\varphi)$ and the symplectic-matrix as $\omega^{ab}$.
The equations of motions are then:

\begin{equation}
{\dot\varphi }^{a}=\omega^{ab}{\partial H\over\partial\varphi ^{b}} \label{uno}
\end{equation} \vspace{0 mm} $\!\!\!$

We shall now {\it suggest}, as path integral for CM, one that forces all paths in ${\cal M}$
to sit on the classical ones. The {\it classical} analog~
$Z_{\scriptscriptstyle CM}$  of the quantum generating 
functional  would be:

\begin{equation}
Z_{\scriptscriptstyle CM}[J]=N\int~{\cal D}\varphi ~{\tilde{\delta}}[\varphi (t)-\varphi _{cl}(t)]~exp\int 
~J\varphi \;dt \label{due}
\end{equation} \vspace{0 mm} \noindent

where $\varphi$ are the $\varphi^{a}\in{\cal M}$, $\varphi_{cl}$ are the solutions of eq.(\ref{uno}),
J is an external current and $\widetilde{\delta}[\;\;]$ is a functional Dirac-delta which forces
every path $\varphi(t)$ to sit on a classical one $\varphi_{cl}(t)$. Of course there are all
possible initial conditions integrated over in (\ref{due}).
One should be very careful in properly defining the measure and the functional
Dirac delta. This careful analysis  has been done
in the literature~\cite{Jona} for stochastic evolution equations and it 
applies  to  Hamiltonian deterministic equations as well.

We should now check if the path integral of eq.~(\ref{due}) leads to the well known operatorial 
formulation~\cite{Koop} of CM done via the Liouville operator and the Lie 
derivative. To do that let us first
rewrite the functional Dirac delta in (\ref{due}) as:

\begin{equation}
\label{eq:tre}
{\tilde\delta}[\varphi -\varphi _{cl}]={\tilde\delta}[{\dot\varphi ^{a}-\omega^{ab}
\partial_{b}H]~det [\delta^{a}_{b}\partial_{t}-\omega^{ac}\partial_{c}\partial
_{b}H}]
\end{equation} \vspace{0 mm} \noindent

where we have used the analog of the relation

\begin{equation}
\displaystyle
\delta[f(x)]=\frac{\delta[x-x_i]}{\Bigm|\frac{\partial f}{\partial x}\Bigm|_{x_i}}
\end{equation} \vspace{0 mm} \noindent

The determinant which appears in (\ref{eq:tre}) is always positive and so we can drop 
the modulus sign $|\;\;|$. 
The next step is to insert (\ref{eq:tre}) in (\ref{due}) and write the $\tilde{\delta}[\;\;]$
as a Fourier transform over some new variables $\lambda_{a}$, i.e.:

\begin{equation}
\label{eq:quattro}
{\tilde{\delta}}\biggl[{\dot\varphi }^{a}-\omega^{ab}{\partial H\over\partial\varphi ^{b}}\biggr]=
\int~{\cal D}\lambda_{a}~exp~i\int\lambda_{a}\biggl[{\dot \varphi }^{a}-\omega^{ab}
{\partial H\over\partial\varphi ^{b}}\biggr]dt
\end{equation} \vspace{0 mm} \noindent

and to re-write the determinant $det[\delta^{a}_b\partial_t-\omega^{ac}\partial_c\partial_bH]$ via
Grassmannian variables $\bar{c}_a, c^{a}$:

\begin{equation}
\label{eq:cinque}
det[\delta^{a}_{b}\partial_{t}-\omega^{ac}\partial_{c}\partial_{b}H]
=\int~{\cal D}c^{a}{\cal D}{\bar c}_{a}~exp~-\int {\bar c}_{a}[\delta^{a}_{b}
\partial_{t}-\omega^{ac}\partial_{c}\partial_{b}H]c^{b}~dt
\end{equation} \vspace{0 mm} \noindent

Inserting (\ref{eq:quattro}), (\ref{eq:cinque}) and (\ref{eq:tre}) 
in (\ref{due}) we get:

\begin{equation}
\label{eq:sei}
Z_{\scriptscriptstyle CM}[0]=\int~{\cal D}\varphi ^{a}{\cal D}\lambda_{a}{\cal D}c^{a}{\cal D}
{\bar c}_{a}~exp~\biggl[i\int~dt~{\widetilde{\cal L}}\biggr]
\end{equation} \vspace{0 mm} \noindent

where $\widetilde{\cal L}$ is:

\begin{equation}
\label{eq:sette}
{\widetilde{\cal L}}=\lambda_{a}[{\dot\varphi }^{a}-\omega^{ab}\partial_{b}H]+
i{\bar c}_{a}[\delta^{a}_{b}\partial_{t}-\omega^{ac}\partial_{c}\partial_{b}H]
c^{b}
\end{equation} \vspace{0 mm} \noindent

One can easily see that this Lagrangian gives the following equations of motion:

\begin{eqnarray}
\label{eq:otto}
{\dot\varphi }^{a}-\omega^{ab}\partial_{b}H & = & 0 \\
\label{eq:nove}
[\delta^{a}_{b}\partial_{t}-\omega^{ac}\partial_{c}\partial_{b}H]c^{b}
 & = & 0 \\
\label{eq:dieci}
\delta^{a}_{b}\partial_{t}{\bar c}_{a}+{\bar
c}_{a}\omega^{ac}\partial_{c}\partial_{b}H & = & 0 \\
\label{eq:undici}
[\delta_{b}^{a}\partial_{t}+\omega^{ac}\partial_{c}\partial_{b}H]\lambda_{a}
& = & -i{\bar c}_{a}\omega^{ac}\partial_{c}\partial_{d}\partial_{b}H c^{d}
\end{eqnarray} \vspace{0 mm} \noindent

One notices immediately the following two things:\\
1) $\widetilde{\cal L}$ leads to the same Hamiltonian equations for $\varphi$ as H did;\\
2) $c^b$ transforms under the Hamiltonian vector field $h\equiv\omega^{ab}
\partial_bH\partial_{a}$
as a {\it form} $d\varphi^{b}$ does.

From the above formalism one can get the equations of motion (\ref{eq:otto})-(\ref{eq:undici})
also via an Hamiltonian $\widetilde{\cal H}$:

\begin{equation}
\label{accatilde}
\widetilde{\cal H}=\lambda_a\omega^{ab}\partial_bH+i\bar{c}_a\omega^{ac}(\partial_c\partial_bH)c^b
\end{equation} \vspace{0 mm} \noindent

and via some
extended Poisson brackets ({\it EPB}) defined in the space \quattrova. 
They are:

\begin{equation}
\label{extPoisbrack}
\{\varphi ^{a},\lambda_{b}\}_{\scriptscriptstyle EPB}=\delta^{a}_{b}~~;~~\{{\bar c}_{b},
c^{a}\}_{\scriptscriptstyle EPB}=-i\delta^{a}_{b} 
\end{equation} \vspace{0 mm} \noindent

The equations of motion (\ref{eq:otto})-(\ref{eq:undici}) are then reproduced 
via the formula 
$\frac{d}{dt}A=\{A,\widetilde{\cal H}\}_{\scriptscriptstyle EPB}$
where A is one of the variables \quattrova.
All the other {\it EPB} are zero; in particular $\{\varphi^{a},\varphi^{b}\}_
{\scriptscriptstyle EPB}=0$ and so these are not the standard Poisson
brackets on ${\cal M}$ which would give $\{\varphi^{a},\varphi^{b}\}_
{\scriptscriptstyle PB}=\omega^{ab}$.

Being (\ref{eq:sei}) a path integral one could also introduce the concept of {\it commutator}
as Feynman did in the quantum case. If we define the graded commutator of two functions $O_1(t)$ and
$O_2(t)$ as the expectation value 
$\langle\;\;\;\rangle$ under our path integral of some time-splitting combinations of the functions
themselves, as:

\begin{equation}
\label{eq:tredici}
\langle[O_{1}(t),O_{2}(t)]\rangle\equiv  \lim_{\epsilon\rightarrow 0}
\langle O_{1}(t+\epsilon)O_{2}(t)\pm O_{2}(t+\epsilon)O_{1}(t)\rangle 
\end{equation} \vspace{0 mm} \noindent

then we get immediately from (\ref{eq:sei}) that the only expressions 
different from zero are:

\begin{equation}
\label{eq:quattordici}
\langle[\varphi ^{a},\lambda_{b}]\rangle=i\delta^{a}_{b}~~;~~\langle[{\bar c}_{b},
c^{a}]\rangle=\delta^{a}_{b}
\end{equation} \vspace{0 mm} \noindent

We notice immediately two things:\\
1) there is an isomorphism between the extended Poisson structure (\ref{extPoisbrack}) and
the graded commutator structure (\ref{eq:quattordici}):
$\{\cdot,\cdot\}_{\scriptscriptstyle EPB}\longrightarrow -i[\cdot,\cdot]$;\\
2) via the commutator structure (\ref{eq:quattordici}) one can "realize" $\lambda_a$ and $\bar{c}_a$
as:

\begin{equation}
\label{eq:sedici}
\lambda_{a}=-i{\partial\over\partial\varphi ^{a}}~~;~~{\bar c}_{a}={\partial\over
\partial c^{a}}
\end{equation} \vspace{0 mm} \noindent

Now, using (\ref{eq:sedici}), we can check that actually what we got as weight in (\ref{eq:sei})
corresponds to the operatorial version of CM. In fact take, for the moment, only the bosonic
(B) part of $\widetilde{\cal H}$ in (\ref{accatilde}):

\begin{equation}
\label{eq:diciassette}
{\widetilde{\cal H}}_{\scriptscriptstyle B}=\lambda_{a}\omega^{ab}\partial_{b}H
\end{equation} \vspace{0 mm} \noindent

This one, via (\ref{eq:sedici}), goes into the operator:

\begin{equation}
\label{eq:diciotto}
{\widehat{\widetilde{\cal H}}}_{\scriptscriptstyle B}\equiv -i\omega^{ab}\partial_{b}H\partial_{a}
\end{equation} \vspace{0 mm} \noindent

which is the Liouville operator of CM. If we had added the Grassmannian part to 
${\widetilde{\cal H}}_{\scriptscriptstyle B}$ and inserted the operatorial representation of 
$\bar{c}$ (\ref{eq:sedici}), we would have got the Lie derivative of the
Hamiltonian flow as we will see in the next section.
So this proves that the operatorial version of CM comes from  
a path-integral weight that is just a Dirac delta on 
the classical paths. Somehow this is the {\it classical}
anologue of what Feynman did for {\it Quantum}
Mechanics where he proved that the Schroedinger operator of evolution
comes from a path-integral weight of the form $exp\,i S$.

\section{Cartan calculus}

We have seen in Sec. 2 that $c^{a}$ transform as  $d\varphi^{a}$, that is as 
the {\it basis} of generic forms $\alpha\equiv \alpha_{a}(\varphi)d\varphi^{a}$ or as the
{\it components} of {\it tangent} vectors :
$V^{a}(\varphi){\partial\over\partial\phi^{a}}$.
The space whose coordinates are $(\varphi^{a}, c^{a})$ is called~\cite{Schw} 
{\it reversed-parity tangent bundle}, indicated as $\Pi T{\cal M}$. The 
"{\it reversed-parity}" specification
is because the $c^{a}$ are Grassmannian variables. As the $(\lambda_{a},
{\bar c}_{a})$ are the "momenta" of the previous variables 
(see eq. (\ref{eq:diciotto})),
we conclude that the 8n variables \quattrova~span the cotangent bundle to the
reversed-parity tangent bundle~\cite{Regini}:~$T^{\star}(\Pi T{\cal M})$.
For more details about this we refer the interested reader to ref.~\cite{Regini}.
So our space is a cotangent bundle and this is the reason why it has a Poisson
structure which is the one we found via the CPI and indicated in
eq. (\ref{eq:tredici}).

In the remaining part of this section we will show how to reproduce all the abstract Cartan calculus
via our {\it EPB} and the Grassmannian variables. Let us first introduce 5 charges which are
conserved under the $\widetilde{\cal H}$ of eq. (\ref{accatilde}) and which 
will play an important role in the Cartan calculus. They are:

\begin{eqnarray}
\label{eq:ventuno}
Q^{\scriptscriptstyle BRS} & \equiv & i c^{a}\lambda_{a} \\
\label{eq:ventidue}
{\bar Q}^{\scriptscriptstyle BRS} & \equiv & i {\bar c}_{a}\omega^{ab}\lambda_{b} \\
\label{eq:ventitre}
Q_{g} & \equiv & c^{a}{\bar c}_{a} \\
\label{eq:ventiquattro}
K & \equiv & {1\over 2}\omega_{ab}c^{a}c^{b} \\
\label {eq:venticinque}
{\bar K} & \equiv & {1\over 2}\omega^{ab}{\bar c}_{a}{\bar c}_{b}
\end{eqnarray} \vspace{0 mm} \noindent

where $\omega_{ab}$ are the matrix elements of the inverse of $\omega^{ab}$.
The next thing we should observe is that $\bar{c}_a$
transforms under the Hamiltonian flow as the basis of  vector fields, 
see eq. (\ref{eq:dieci}),
while $\lambda_{a}$
does not seem to transform as a vector field, eq. (\ref{eq:undici}), even if it can be 
interpreted as $\frac{\partial}{\partial \varphi^{a}}$. The
explanation of this fact is given in ref.~\cite{Regini}.

Now since $c^{a}$ transforms as basis of forms $d\varphi^{a}$ and $\bar{c}_a$ 
as basis of vector fields\footnote{This is so not only under the Hamiltonian
flow but under any diffeomorphism: see ref. \cite{Regini} for details.} $\frac{\partial}
{\partial \varphi^{a}}$, let us start building the following map, called "hat"
map $\wedge$:

\begin{eqnarray}
\label{eq:ventisei}
\alpha=\alpha_{a}d\varphi ^{a} & \hat{\longrightarrow} &  {\widehat\alpha}\equiv
\alpha_{a}c^{a}\\
\label{eq:ventisette}
V=V^{a}\partial_{a}  & \hat{\longrightarrow} & {\widehat V}\equiv V^{a}{\bar
c}_{a}
\end{eqnarray} \vspace{0 mm} \noindent

It is  actually a much more general map between forms $\alpha$, antisymmetric
tensors $V$ 
and functions of $\varphi, c, \bar{c}$:

\begin{eqnarray}
\label{eq:ventotto}
F^{(p)}={1\over p !}F_{a_{1}\cdots a_{p}}d\varphi ^{a_{1}}\wedge\cdots\wedge
d\varphi ^{a_{p}}  & \hat{\longrightarrow} &{\widehat F}^{(p)}\equiv {1\over p!}
F_{a_{1}\cdots a_{p}}c^{a_{1}}\cdots c^{a_{p}}\\
\label{eq:ventinove}
V^{(p)}={1\over p!}V^{a_{1}\cdots a_{p}}\partial_{a_{1}}\wedge\cdots\wedge 
\partial_{a_{p}} & \hat{\longrightarrow} & {\widehat V}\equiv {1\over p!}V^{a_{1}
\cdots a_{p}}{\bar c}_{a_{1}}\cdots {\bar c}_{a_{p}}
\end{eqnarray} \vspace{0 mm} \noindent

Once the correspondence (\ref{eq:ventisei})-(\ref{eq:ventinove}) is extablished 
we can easily find  out what correspond to the various Cartan operations 
known as the exterior derivative {\bf d}
of a form, or interior contraction between a vector field $V$ and a form $F$. It
is easy to check that, see \cite{Gozzi}:

\begin{eqnarray}
\label{eq:trenta}
dF^{(p)} & \hat{\longrightarrow} & i\{Q^{\scriptscriptstyle BRS},{\widehat F}^{(p)}\}_{\scriptscriptstyle EPB} \\
\label{eq:trentuno}
\iota_{{\scriptscriptstyle V}}F^{(p)} & \hat{\longrightarrow} & i\{{\widehat V},
{\widehat F}^{(p)}\}_{\scriptscriptstyle EPB}
\\
\label{eq:trentadue}
pF^{(p)} & \hat{\longrightarrow} & i\{Q_{g}, {\widehat F}^{(p)}\}_{\scriptscriptstyle EPB}
\end{eqnarray} \vspace{0 mm} \noindent

where $Q^{\scriptscriptstyle BRS}, \, Q_g$ are the charges of (\ref{eq:ventuno})-(\ref{eq:ventitre}).
At the same level we can translate in our language the usual mapping~\cite{Marsd}
between vector fields $V$
and forms $V^{\flat}$ realized by the symplectic 2-form $\omega(V,0)\equiv V^{\flat}$,
or the inverse operation of building a vector field $\alpha^{\sharp}$ out of a form:
$\alpha=(\alpha^{\sharp})^{\flat}$. These operations can be translated in our
formalism as follows:

\begin{eqnarray}
V^{\flat} & \hat{\longrightarrow} & i\{K,{\widehat V}\}_{\scriptscriptstyle EPB}\\
\label{eq:trentaquattro}
\alpha^{\sharp} & \hat{\longrightarrow} & i\{{\bar
K},{\widehat\alpha}\}_{\scriptscriptstyle EPB}
\end{eqnarray} \vspace{0 mm} \noindent

where again $K, \bar{K}$ are the charges (\ref{eq:ventiquattro})-(\ref{eq:venticinque}).
We can also translate the standard operation of building a vector field out of a function 
$f(\varphi)$, and also the Poisson brackets between two functions $f$ and $g$:

\begin{eqnarray}
\label{eq:trantacinque}
(df)^{\sharp} & \hat{\longrightarrow} & i\{{\bar Q}^{\scriptscriptstyle BRS},f\}_{\scriptscriptstyle EPB}\\
\label{eq:trentasei}
\{f,g\}_{\scriptscriptstyle PB}=df[(dg)^{\sharp}] & \hat{\longrightarrow} & 
-\{\{\{f,Q^{\scriptscriptstyle BRS}\},{\bar K}\},
\{\{\{g,Q^{\scriptscriptstyle BRS}\},{\bar K}\},K\}\}_{\scriptscriptstyle EPB}
\end{eqnarray} \vspace{0 mm} \noindent
 
The next thing to do is to translate the concept of Lie derivative which 
is defined as:  
~${\cal L}_{\scriptscriptstyle V}=d\iota_{\scriptscriptstyle V}
+\iota_{\scriptscriptstyle V}d$. It is easy to prove that:

\begin{equation}
\label{eq:trentasette}
{\cal L}_{\scriptscriptstyle V}F^{(p)} \;\; \hat{\longrightarrow} \;\; 
\{-{\widetilde {\cal H}}_{\scriptscriptstyle V},{\widehat F}
^{(p)}\}_{\scriptscriptstyle EPB}
\end{equation} \vspace{0 mm} \noindent

where ${\widetilde {\cal H}}_{\scriptscriptstyle V}=\lambda_aV^{a}+i\bar{c}_a
\partial_bV^{a}c^{b}$; note that,
for $V^{a}=\omega^{ab}\partial_bH$,  ${\widetilde {\cal H}}_{\scriptscriptstyle V}$ becomes the ${\widetilde 
{\cal H}}$
of (\ref{accatilde}). This confirms that the full ${\widetilde {\cal H}}$ of eq. (\ref{accatilde})
is the Lie derivative of the Hamiltonian flow as we said at the end of the previous section. Finally
the Lie brackets between two vector fields $V,\;W$ are reproduced by:

\begin{equation}
\label{eq:trentotto}
[V,W]_{Lie-brack.} \; \hat{\longrightarrow} \; \{-{\widetilde {\cal H}}_{\scriptscriptstyle 
V},{\widehat W}\}_{\scriptscriptstyle EPB}
\end{equation} \vspace{0 mm} \noindent

We will collect in Appendix A all the important formulas we mentioned in this section. 

\section{Schouten-Nijenhuis, Fr\"olicher-Nijenhuis and\break
Nijenhuis-Richardson brackets done via the\break extended Poisson brackets}

\subsection{Schouten-Nijenhuis (SN) brackets}

These brackets are a generalization of the Lie brackets between vector fields: 
in fact the SN are brackets
between {\it multivector fields}  and they become the usual Lie 
brackets in case of vector fields.
As Lie brackets associate to two vector fields $X$ and $Y$ another vector field $[X,Y]$, so SN 
brackets associate to two multivector fields of rank $p$\break ($P=X_{(1)}\wedge\cdots\wedge X_{(p)}$)
and $r$ ($R=Y_{(1)}\wedge\cdots\wedge Y_{(r)}$) a multivector field of rank $p+r-1$ via the
following rule~\cite{vaisman}:

\begin{eqnarray}
&&\qquad\qquad [\cdot,\cdot]_{\scriptscriptstyle SN}: \; {\cal V}^p(M)\times{\cal V}^r(M) \; 
\longrightarrow \; {\cal V}^{p+r-1}(M)
\nonumber\\
&&[P,R]_{\scriptscriptstyle SN}
\equiv \sum_{i=1}^{p}(-1)^{i+1}X_{(1)}\wedge\cdots
\wedge{\widehat {\widehat  X}} _{(i)}\cdots\wedge X_{(p)}\wedge[X_{(i)},R]
\end{eqnarray} \vspace{0 mm} \noindent

where the ${\cal V}^{s}$ indicates the space of mutivector fields of rank
s and the double hat ${\widehat {\widehat  X}}_{(i)}$ indicates that we have removed
 the $X_{(i)}$, while 
$[X_{(i)},R]={\cal L}_{\scriptscriptstyle X_{(i)}}R$ is the 
Lie derivative of a multivector defined as:

\begin{equation}
{\cal L}_{{\scriptscriptstyle X}_{(i)}}R=
\sum_{j=1}^rY_1\wedge\cdots\wedge[X_{(i)},Y_{(j)}]
\wedge\cdots\wedge Y_{(r)}
\end{equation} \vspace{0 mm} \noindent

This Lie derivative is reproduced via our extended Poisson brackets 
({\it EPB}) as:

\begin{equation}
{\cal L}_{\scriptscriptstyle X_{(i)}}R \;\; \hat{\longrightarrow} \;\; 
\{-\widetilde{\cal H}_{\scriptscriptstyle X_{(i)}}, \widehat{R}\}_{\scriptscriptstyle EPB}
\end{equation} \vspace{0 mm} \noindent

where we have  defined: $\widetilde{\cal H}_{\scriptscriptstyle X_{(i)}}=\{\widehat{X}_{(i)}, Q
^{\scriptscriptstyle BRS}\}_{
\scriptscriptstyle EPB}$.
The SN brackets become then in our notation :

\begin{equation}
[P,R]_{\scriptscriptstyle SN} \;\; \hat{\longrightarrow} \;\; 
-\{\widetilde{\cal H}_{\scriptscriptstyle P}, \widehat {R}\}_
{\scriptscriptstyle EPB}
\end{equation} \vspace{0 mm} \noindent

where:

\begin{equation}
\widetilde{\cal H}_{\scriptscriptstyle P}=\{Q,\widehat{X}_{(1)}\cdots\widehat{X}_{(p)}\}
=\sum_{i=1}^p(-1)^{i+1}\widehat{X}_{(1)}\cdots{\widehat{\widehat X}}_{(i)}\cdots\widehat{X}_{(p)}
\widetilde{\cal H}_{\scriptscriptstyle X_{(i)}}
\end{equation} \vspace{0 mm} \noindent

and:

\begin{equation}
\widehat{R}=Y_{(1)}^{j_1}\bar{c}_{j_1}\cdots
Y_{(r)}^{j_r}\bar{c}_{j_r}
\end{equation} \vspace{0 mm} \noindent

The quantities which one has in the equations above are  those written in terms of $c^{a}$ or
${\bar c}_{a}$ as explained in the previous section.

The extended Poisson brackets~({\it EPB}), besides allowing  us to write complicated 
formulas in a very compact way,
can also be used to prove some properties of the Schouten-Nijenhuis brackets, as we will show
in Appendix B.

\subsection{Fr\"olicher-Nijenhuis (FN) Brackets}

These are brackets which associate to two {\it vector-valued forms}\footnote{The
notation we follow, regarding the manner to
indicate the space of vector valued forms with $\Omega^{k}(M,TM)$, is
the one of ref.~\cite{Kolar}.} $K\in\Omega^{k+1}(M;TM)$ 
of degree $k+1$ and $L\in\Omega^{l+1}(M;TM)$ of degree $l+1$ 
a vector-valued form of degree $k+l+2$:

\begin{equation}
[\cdot,\cdot]_{\scriptscriptstyle FN}: \; \Omega^{k+1}(M;TM)\times\Omega^{l+1}(M;TM)
\; \longrightarrow \; \Omega^{k+l+2}(M;TM)
\end{equation} \vspace{0 mm} \noindent

They are defined in the following manner~\cite{Kolar}:

a) let us first define the interior contraction $\iota_{\scriptscriptstyle K}$ 
not with a vector field but with a  
vector-valued form $K$ of degree $k+1$, and apply it on a form $\omega$ of degree $l$. As $K$
is a $(k+1)$-form, $\iota_{\scriptscriptstyle K}\omega$ can eat $k+l$ vector fields, i.e. 
when we apply $\iota_{\scriptscriptstyle K}\omega$ to $k+l$ vectors, 
we obtain the following number:

\begin{eqnarray}
&&\qquad \qquad(\iota_{{\scriptscriptstyle K}}\omega)(X_{(1)},\cdots,X_{(k+l)}) \equiv \nonumber\\
&&\equiv {1\over (k+1)!(l-1)!} \sum_{\{\sigma\in S_{k+l}\}} 
(sign~\sigma)\;
\omega [K(X_{\sigma(1)},\cdots,X_{\sigma(k+1)}), X_{\sigma(k+2)},\cdots, X_{\sigma(k+l)}] \nonumber\\
&&
\end{eqnarray} \vspace{0 mm} \noindent

where $S_{k+l}$ is the set of permutations of the $k+l$ vector fields $X_{(1)}\cdots X_{(k+l)}$. We
note how the $k+1$ vector-valued form $K$, acting on $k+1$ vector fields, produces another vector 
field;

b) having defined this generalized interior product $\iota_{\scriptscriptstyle K}$, we can also define a 
generalized
Lie derivative as: 

\begin{equation}
{\cal L}_{\scriptscriptstyle K} \; = \; [\iota_{\scriptscriptstyle K}, d]
\end{equation} \vspace{0 mm} \noindent

where $[\cdot , \cdot]$ is the usual graded commutator and $K\in \Omega^{k+1}(M;TM)$.

c) Now, having done a) and b), the FN brackets are defined in the 
following implicit way:

\begin{equation}
[{\cal L}_{\scriptscriptstyle K},{\cal L}_{\scriptscriptstyle L}] \; \equiv \;
{\cal L}_{\scriptscriptstyle [K,L]_{\scriptscriptstyle FN}}
\end{equation} \vspace{0 mm} \noindent

where $[{\cal L}_{\scriptscriptstyle K}, {\cal L}_{\scriptscriptstyle L}]$  
is the usual graded commutator among Lie derivatives.

Now if $K$ and $L$ are written in our linguage as:

\begin{eqnarray}
\displaystyle
K & \hat{\longrightarrow} & \frac{1}{(k+1)!}\;K^{i}_{i_1i_2\cdots i_{k+1}}\;[c^{i_1}c^{i_2}\cdots c^{i_{k+1}}]
[\bar{c}_i] \nonumber\\
L & \hat{\longrightarrow} & \frac{1}{(l+1)!}\;\;\,L^{j}_{j_1j_2\cdots j_{l+1}}\;[c^{j_1}c^{j_2}\cdots c^{j_{l+1}}]
[\bar{c}_j] \label{cappa&elle}
\end{eqnarray} \vspace{0 mm} \noindent

then the FN brackets become:

\begin{equation}
[K,L]_{\scriptscriptstyle FN} \; \hat{\longrightarrow} \; -
\{\widetilde{\cal H}_{\scriptscriptstyle K},\widehat{L}\}_{\scriptscriptstyle EPB}
\end{equation} \vspace{0 mm} \noindent

where:

\begin{equation}
\displaystyle
\widetilde{\cal H}_{\scriptscriptstyle K}=\frac{1}{(k+1)!}\Biggl(\lambda_jK^j_{j_1j_2\cdots j_{k+1}}
+i\bar{c}_j(\partial_dK^{j}_{j_1j_2\cdots j_{k+1}}c^d)\Biggr)c^{j_1}\cdots c^{j_{k+1}}
\end{equation} \vspace{0 mm} \noindent

The calculational details are given in Appendix C.
\subsection{Nijenhuis-Richardson (NR) brackets}

They are brackets defined among two vector-valued forms: $K\in\Omega^{k+1}(M;TM)$ and $L\in\Omega
^{l+1}(M;TM)$ and they give a $k+l+1$ vector-valued form defined in the 
following implicit way:

\begin{eqnarray}
&&[\cdot,\cdot]_{\scriptscriptstyle NR}: \; \Omega^{k+1}(M;TM)\times\Omega^{l+1}(M;TM) \;
\longrightarrow \; \Omega^{k+l+1}(M;TM)\nonumber\\
&&\qquad\qquad\qquad\qquad\iota_{\scriptscriptstyle [K,L]_{\scriptscriptstyle NR}}\; \equiv 
\;[\iota_{\scriptscriptstyle K},
\iota_{\scriptscriptstyle L}] \label{defNR1}
\end{eqnarray} \vspace{0 mm} \noindent

where $\iota_{\scriptscriptstyle K}$ and $\iota_{\scriptscriptstyle L}$ are the generalized
interior contractions defined in the previous section. The definition (\ref{defNR1}) can also
be expressed in a more explicit way as:

\begin{equation}
[K,L]_{\scriptscriptstyle NR}\; = \;\iota_{\scriptscriptstyle K}L-(-1)^{kl}
\iota_{\scriptscriptstyle L}K
\end{equation} \vspace{0 mm} \noindent

Now with our calculus and $\widehat{K}$ and $\widehat{L}$ defined as in 
(\ref{cappa&elle}) we can 
show that :

\begin{equation}
[K,L]_{\scriptscriptstyle NR} \;\; \hat{\longrightarrow} \;\; i\{\widehat{K},\widehat{L}\}_{\scriptscriptstyle
ePb}
\end{equation} \vspace{0 mm} \noindent

The calculational details are provided in  Appendix D.

We can now summarize all SN, FN, NR brackets in the following very compact way:

\begin{eqnarray}
\label{centrale}
&&[P,R]_{\scriptscriptstyle SN} \;\; \hat{\longrightarrow} \;\; -\{
\widetilde{\cal H}_{\scriptscriptstyle P},\widehat {R}\}_{\scriptscriptstyle EPB}\nonumber\\
&&[K,L]_{\scriptscriptstyle FN} \;\; \hat{\longrightarrow} \;\; 
-\{\widetilde{\cal H}_{\scriptscriptstyle K}, \widehat{L}\}_{\scriptscriptstyle EPB}\nonumber\\
&&[K,L]_{\scriptscriptstyle NR} \;\; \hat{\longrightarrow} \;\;\;\;\; i\{\widehat{K},\widehat{L}\}_{\scriptscriptstyle
EPB}
\end{eqnarray} \vspace{0 mm} \noindent

where:

\begin{eqnarray}
&&R=Y_{(1)}\wedge\cdots\wedge Y_{(r)} \;\;\;\;\, \hat{\longrightarrow} \;\;\; Y_{(1)}^{j_1}\bar{c}_{j_1}\cdots
Y_{(r)}^{j_r}\bar{c}_{j_r}\nonumber\\
&&\widetilde{\cal H}_{\scriptscriptstyle P}=\{Q,\widehat{X}_{(1)}\cdots\widehat{X}_{(p)}\}
=\sum_{i=1}^p(-1)^{i+1}\widehat{X}_{(1)}\cdots{\widehat{\widehat X}}_{(i)}\cdots\widehat{X}_{(p)}
\widetilde{\cal H}_{\scriptscriptstyle X_{(i)}}\nonumber\\
&&\displaystyle
\widetilde{\cal H}_{\scriptscriptstyle K}=\frac{1}{(k+1)!}\Biggl(\lambda_jK^j_{j_1j_2\cdots j_{k+1}}
+i\bar{c}_j(\partial_dK^{j}_{j_1j_2\cdots j_{k+1}}c^d)\Biggr)c^{j_1}\cdots c^{j_{k+1}}\nonumber\\
&&K\;\in\Omega^{k+1}(M;TM) \;\;\;\;\; \hat{\longrightarrow} \;\;  
\frac{1}{(k+1)!}K^{i}_{i_1i_2\cdots i_{k+1}}[c^{i_1}c^{i_2}\cdots c^{i_{k+1}}]
[\bar{c}_i] \nonumber\\
&&L\;\;\in\Omega^{l+1}(M;TM) \;\;\;\;\;\, \hat{\longrightarrow} \;\;
\frac{1}{(l+1)!}\;\,L^{j}_{j_1j_2\cdots j_{l+1}}[c^{j_1}c^{j_2}\cdots c^{j_{l+1}}]
[\bar{c}_j]
\end{eqnarray} \vspace{0 mm} \noindent

\section{Conclusions}
The reader may ask which is the use of all this. Our answer is that  by looking
at eq. (\ref{centrale}) one immediately realizes that we  have
reduced {\it three} different and complicated brackets, like the SN, FN
and NR brackets, to only {\it one} bracket which is  our 
{\bf E}xtended {\bf P}oisson {\bf B}racket (or {\it EPB}) in which
the entries are different functions of our variables. So instead of changing the
brackets we just have to change the  entries to reproduce all the three
SN, FN and NR brackets. This unifying structure is not only appealing
but it may also indicate something more profound which may be worth
to investigate in the future.

\newpage

\section*{APPENDIX A}
{\centerline {\bf CARTAN CALCULUS}
The correspondence between the standard Cartan calculus~\cite{Marsd} and our
formulation is provided below:

\begin{eqnarray}
\label{eq:A-uno}
Q^{\scriptscriptstyle BRS}  \equiv  i c^{a}\lambda_{a}&;&{\bar Q}^{\scriptscriptstyle BRS}  \equiv  
i {\bar c}_{a}\omega^{ab}\lambda_{b} \\
\label{eq:A-due}
Q_{g}  &\equiv & c^{a}{\bar c}_{a} \\
\label{eq:a-duebis} 
K \equiv  {1\over 2}\omega_{ab}c^{a}c^{b} &;& 
{\bar K} \equiv {1\over 2}\omega^{ab}{\bar c}_{a}{\bar c}_{b}\\
\label{eq:A-tre}
\{\varphi ^{a},\lambda_{b}\}_{\scriptscriptstyle EPB}=\delta^{a}_{b}&;&\{{\bar
c}_{b},c^{a}\}_{\scriptscriptstyle EPB}=-i\delta^{a}_{b}\\
\label{eq:A-quattro}
\alpha=\alpha_{a}d\varphi ^{a} & \hat{\longrightarrow} &  {\widehat\alpha}\equiv
\alpha_{a}c^{a}\\
\label{eq:A-cinque}
V=V^{a}\partial_{a}  & \hat{\longrightarrow} & {\widehat V}\equiv V^{a}{\bar
c}_{a}\\
\label{eq:A-sei}
F^{(p)}={1\over p !}F_{a_{1}\cdots a_{p}}d\varphi ^{a_{1}}\wedge\cdots\wedge
d\varphi ^{a_{p}}  & \hat{\longrightarrow} &{\widehat F}^{(p)}\equiv {1\over p!}
F_{a_{1}\cdots a_{p}}c^{a_{1}}\cdots c^{a_{p}}\\
\label{eq:A-sette}
V^{(p)}={1\over p!}V^{a_{1}\cdots a_{p}}\partial_{a_{1}}\wedge\cdots\wedge 
\partial_{a_{p}} & \hat{\longrightarrow} & {\widehat V}\equiv {1\over p!}V^{a_{1}
\cdots a_{p}}{\bar c}_{a_{1}}\cdots {\bar c}_{a_{p}}\\
\label{eq:A-otto}
dF^{(p)} & \hat{\longrightarrow} & i\{Q^{\scriptscriptstyle BRS},{\widehat F}^{(p)}\}_{\scriptscriptstyle EPB} \\
\label{eq:A-nove}
\iota_{\scriptscriptstyle V}F^{(p)} & \hat{\longrightarrow} & i\{{\widehat V},
{\widehat F}^{(p)}\}_{\scriptscriptstyle EPB}
\\
\label{eq:A-dieci}
pF^{(p)} & \hat{\longrightarrow} & i\{Q_{g}, {\widehat F}^{(p)}\}_{\scriptscriptstyle EPB}\\
\label{eq:A-undici}
V^{\flat} & \hat{\longrightarrow} & i\{K,{\widehat V}\}_{\scriptscriptstyle EPB}\\
\label{eq:A-dodici}
\alpha^{\sharp} & \hat{\longrightarrow} & i\{{\bar
K},{\widehat\alpha}\}_{\scriptscriptstyle EPB}\\
\label{eq:A-tredici}
(df)^{\sharp} & \hat{\longrightarrow} & i\{{\bar Q}^{\scriptscriptstyle BRS},f\}_{\scriptscriptstyle EPB}\\
\label{eq:A-quattordici}
\{f,g\}_{\scriptscriptstyle PB}=df[(dg)^{\sharp}] & \hat{\longrightarrow} & -\underbrace{\{\{\{}
_{\scriptscriptstyle EPB} f,Q^{\scriptscriptstyle BRS}\},{\bar
K}\},\underbrace{\{\{\{}_{\scriptscriptstyle EPB} g,Q^{\scriptscriptstyle BRS}\},{\bar K}\},K\}\}
\nonumber\\
&&\\
\label{eq:A-quindici}
{\cal L}_{\scriptscriptstyle V}F^{(p)} & \hat{\longrightarrow} & -\{{\widetilde {\cal H}}_
{\scriptscriptstyle V},{\widehat F}
^{(p)}\}_{\scriptscriptstyle EPB}\\
\label{eq:A-sedici}
[V,W]_{Lie-brack.} & \hat{\longrightarrow} & -\{{\widetilde {\cal H}}_{\scriptscriptstyle 
V},{\widehat
W}\}_{\scriptscriptstyle EPB}
\end{eqnarray} \vspace{0 mm} \noindent

\newpage
\section*{APPENDIX B} 
{\centerline {\bf CALCULATIONAL DETAILS REGARDING THE SN BRACKETS}}

We report here some more detailed calculations about the SN brackets and we 
follow the book \cite{vaisman}.
From now on all the curly brackets  mean {\it EPB}-brackets and $Q$ indicates
 the  BRS charge that previously we indicated as $Q^{\scriptscriptstyle BRS}$.
Since we will use it widely, we want first to return to the formula regarding
the  Lie brackets
between two vector fields, of which the SN brackets are only a generalization.
As we have seen in (\ref{eq:A-sedici}) the correct translation in our language
of the Lie brackets is 

\begin{equation}
[V,W]_{Lie-brack.} \;\; \hat{\longrightarrow} \;\; -\{{\widetilde {\cal H}}_{\scriptscriptstyle 
V},{\widehat W}\}
\end{equation} \vspace{0 mm} \noindent

In fact:

\begin{eqnarray}
&&-\{{\widetilde {\cal H}}_{\scriptscriptstyle 
V},{\widehat W}\}=-\{\lambda_a V^{a}+i\bar{c}_a(\partial_b V^{a})c^b,W^c\bar{c}_c\}=\nonumber\\
&&=V^{a}\bar{c}_c\partial_aW^c-\bar{c}_a(\partial_bV^{a})W^b=[V^b(\partial_bW^{a})-W^b
(\partial_bV^{a})]\bar{c}_a
\end{eqnarray} \vspace{0 mm} \noindent

So we have obtained, correctly, a vector field whose components are just the 
components of the Lie 
brackets between $V^{a}\bar{c}_a$ and $W^{a}\bar{c}_a$, see \cite{Kolar}.

Next  we extend the concept of {\it interior contraction}: the interior product between a 
p-multivector field $P=X_{(1)}\wedge\cdots\wedge X_{(p)}$ and an 
$l$-form $\omega$ is defined as:

\begin{equation}
\iota_{\scriptscriptstyle P}\omega(\cdots)\;=\;
w(X_{(1)},X_{(2)},\cdots,X_{(p)},\cdots)
\end{equation} \vspace{0 mm} \noindent

Now from the definition itself of interior product we have:

\begin{eqnarray}
&&\iota_{\scriptscriptstyle X_{(p)}}\iota_{\scriptscriptstyle X_{(p-1)}}\cdots
\iota_{\scriptscriptstyle X_{(1)}}\omega(\cdots)=
\iota_{\scriptscriptstyle X_{(p-1)}}\cdots\iota_{\scriptscriptstyle X_{(1)}}\omega(X_{(p)},\cdots)
=\nonumber\\
&&=\iota_{\scriptscriptstyle X_{(p-2)}}\cdots\iota_{\scriptscriptstyle X_{(1)}}\omega(X_{(p-1)},
X_{(p)},\cdots)=
\cdots=\omega(X_{(1)},X_{(2)},\cdots,X_{(p)},\cdots)=\iota_{\scriptscriptstyle P}\omega(\cdots)
\nonumber\\
&&
\end{eqnarray} \vspace{0 mm} \noindent

In this way we can transform the interior product with a multivector into a set of 
interior contractions with normal vector fields:

\begin{equation}
\iota_{\scriptscriptstyle P}\omega=\iota_{\scriptscriptstyle X_{(p)}}
\iota_{\scriptscriptstyle X_{(p-1)}}\cdots \iota_{\scriptscriptstyle X_{(2)}}\iota_
{\scriptscriptstyle X_{(1)}}\omega \label{contraction1}
\end{equation} \vspace{0 mm} \noindent

Let us remember that we know which is the $\wedge$-map of an interior
contraction between a form and a vector:

\begin{equation}
\iota_{\scriptscriptstyle V}\omega \;\; \hat{\longrightarrow} \;\; i\{\widehat{V},\widehat{\omega}\}
\label{contraction2}
\end{equation} \vspace{0 mm} \noindent

So, applying the previous formula over and over again, we obtain from (\ref{contraction1}) and
(\ref{contraction2}):

\begin{eqnarray}
\displaystyle
&&\iota_{\scriptscriptstyle P}\omega \; \hat{\longrightarrow} \; i\{\widehat{X}_{(p)}, 
(\iota_{\scriptscriptstyle X_{(p-1)}}
\cdots\iota_{\scriptscriptstyle X_{(2)}}\iota_{\scriptscriptstyle X_{(1)}}\omega)^{\wedge}\}
=\nonumber\\
&&=i^2\{X^{i_p}_{(p)}\bar{c}_{i_p},\{X^{i_{p-1}}_{(p-1)}\bar{c}_{i_{p-1}},
(\iota_{\scriptscriptstyle X_{(p-2)}}
\cdots\iota_{\scriptscriptstyle X_{(2)}}\iota_{\scriptscriptstyle X_{(1)}}\omega)^{\wedge}\}\}
=\cdots=\nonumber\\
&&=\frac{i^p}{l!}\{X^{i_p}_{(p)}\bar{c}_{i_p},\{X^{i_{p-1}}_{(p-1)}\bar{c}_{i_{p-1}},
\{\cdots,\{X^{i_1}_{(1)}\bar{c}_{i_1},\omega_{ij\cdots l}c^{i}c^j\cdots c^{l}
\underbrace{\}\}\cdots\}}_{\scriptscriptstyle p\;brackets}
\label{intcmultivec}
\end{eqnarray} \vspace{0 mm} \noindent

One could have been tempted to make the $\wedge$-correspondence not with the RHS of
(\ref{intcmultivec}), but with something proportional to:

\begin{equation}
\{X^{i_1}_{(1)}\bar{c}_{i_1}X^{i_2}_{(2)}\bar{c}_{i_2}\cdots X^{i_p}_{(p)}\bar{c}_{i_p},
\omega_{ij\cdots l}c^{i}c^{j}\cdots c^l\} \label{achtung}
\end{equation} \vspace{0 mm} \noindent

but it would be wrong. In fact, while (\ref{intcmultivec}) is an $(l-p)$-form, that is a 
$(l-p)$-string of $c$, (\ref{achtung}) is not a string of only $c$. 

Now we notice that the interior contraction with a 2-vector can be expressed
as a combination of well-known objects, such as Lie-brackets, exterior derivatives 
and interior contractions with normal vector fields. 
In fact, according to our formalism, we have:

\begin{equation}
\label{aggiunta}
\iota_{\scriptscriptstyle V\wedge W}dw \;\; \hat{\longrightarrow} \;\; -i\{\hat{W},\{\hat{V},
\{Q,\hat{\omega}\}\}\}
\end{equation} \vspace{0 mm} \noindent

where $V$ and $W$ are vector fields, $\omega$ is a 1-form and $Q$ is the usual BRS-charge.
Using the Jacobi identity, we can write:

\begin{equation}
\label{aggiuntauno}
-i\{\hat{W},\{\hat{V},\{Q,\hat{\omega}\}\}\}=i\{\hat{W},\{Q,\{\hat{V},\hat{\omega}\}\}\}
+i\{\hat{W},\{\hat{\omega},\{Q,\hat{V}\}\}\}
\end{equation} \vspace{0 mm} \noindent

The last term of (\ref{aggiuntauno}) can be rewritten using again the Jacobi 
identity:

\begin{equation}
\label{aggiuntadue}
+i\{\hat{W},\{\hat{\omega},\{Q,\hat{V}\}\}\}=+i\{\hat{\omega},\{\{Q,\hat{V}\},\hat{W}\}\}
-i\{\{Q,\hat{V}\},\{\hat{W},\hat{\omega}\}\}
\end{equation} \vspace{0 mm} \noindent

Let us now manipulate the last term of (\ref{aggiuntadue}):

\begin{eqnarray}
\label{aggiuntatre}
&&-i\{\{Q,\hat{V}\},\{\hat{W},\hat{\omega}\}\}=+i\{\{\hat{W},\hat{\omega}\},
\{\hat{V},Q\}\}=\nonumber\\
&&-i\{\hat{V},\{Q,\{\hat{W},\hat{\omega}\}\}\}-i\{Q,\{\{\hat{W},\hat{\omega}\},\hat{V}
\}\}
\end{eqnarray} \vspace{0 mm} \noindent

The last term of (\ref{aggiuntatre}) is identically zero, in fact:

\begin{equation}
\{\{\hat{W},\hat{\omega}\},\hat{V}\}=\{\{W^{a}\bar{c}_a,\omega_bc^b\},V^{d}c_d\}
=\{-iW^{a}\omega_a,V^dc_d\}=0
\end{equation} \vspace{0 mm} \noindent

Collecting all the previous results in (\ref{aggiunta}) we obtain:

\begin{eqnarray}
&&\iota_{\scriptscriptstyle V\wedge W}dw \;\; \hat{\longrightarrow} \;\; 
i\{\hat{W},\{Q,\{\hat{V},\hat{\omega}\}\}\}+\nonumber\\
&&+i\{\{\{Q,\hat{V}\},\hat{W}\},\hat{\omega}\}-i\{\hat{V},\{Q,\{\hat{W},\hat{\omega}\}\}\}=\nonumber\\
&&=\biggl(-\iota_{\scriptscriptstyle W}d(\iota_{\scriptscriptstyle V}\omega)
-\iota_{\scriptscriptstyle [V,W]}\omega+\iota_{\scriptscriptstyle V}d(\iota_{\scriptscriptstyle W}
\omega)\biggr)^{\wedge}
\end{eqnarray} \vspace{0 mm} \noindent

So we have obtained the formula:

\begin{equation}
\label{aggiuntaquattro}
\iota_{\scriptscriptstyle V\wedge W}dw=-\iota_{\scriptscriptstyle W}d(\iota_{\scriptscriptstyle V}
\omega)
-\iota_{\scriptscriptstyle [V,W]}\omega+\iota_{\scriptscriptstyle V}d(\iota_{\scriptscriptstyle W}
\omega)
\end{equation} \vspace{0 mm} \noindent

that can be generalized to the case of multivector fields; we can in fact express interior 
contraction with a multivector field using exterior derivatives, interior contractions with
multivectors of lower rank and the SN brackets according to the formula:

\begin{equation}
\label{vaismaniana}
\iota_{\scriptscriptstyle P\wedge R}dw=-\iota_{\scriptscriptstyle R}d(\iota_{\scriptscriptstyle P}
\omega)
-\iota_{\scriptscriptstyle [P,R]_{\scriptscriptstyle SN}}\omega+\iota_{\scriptscriptstyle P}
d(\iota_{\scriptscriptstyle R}\omega)
\end{equation} \vspace{0 mm} \noindent

where $P$ and $R$ are multivector fields of rank p and r respectively and $\omega$ is a\break 
$p+r-1$ form. 
A proof of (\ref{vaismaniana}), that is the natural generalization of (\ref{aggiuntaquattro}),
can be found in \cite{Bhaskara}.

Following \cite{vaisman}, we can define the Schouten-Nijenhuis brackets between two multivector
fields $P=X_{(1)}\wedge X_{(2)}\wedge\cdots\wedge X_{(p)}$ and $R=Y_{(1)}\wedge Y_{(2)}
\wedge\cdots\wedge Y_{(r)}$ as the $(p+r-1)$-multivector field given by:

\begin{equation}
[P,R]_{\scriptscriptstyle SN}=\sum_{i=1}^p(-1)^{i+1}X_{(1)}\wedge\cdots\wedge{\widehat {\widehat  
X}}_{(i)}
\wedge\cdots\wedge X_{(p)}\wedge[X_{(i)},R] \label{SN brackets}
\end{equation} \vspace{0 mm} \noindent

where ${\widehat {\widehat  X}}_{(i)}$ means $X_{(i)}$ is missing and where:

\begin{equation}
[X_{(i)},R]={\cal L}_{\scriptscriptstyle 
X_{(i)}}R=\sum_{j=1}^rY_{(1)}\wedge\cdots\wedge[X_{(i)},Y_{(j)}]\wedge\cdots
\wedge Y_{(r)} \label{liederivative}
\end{equation} \vspace{0 mm} \noindent

In the previous formula $[X_{(i)},Y_{(j)}]$ are the usual Lie brackets between vector fields.
Now (\ref{liederivative}), that is the Lie derivative along a vector field of a 
multivector, can be translated in our language as:

\begin{equation}
{\cal L}_{\scriptscriptstyle X_{(i)}}R \; \hat{\longrightarrow} \; \{-\widetilde{\cal H}_{
\scriptscriptstyle X_{(i)}},
\widehat{R}\}=
-\{\{\widehat{X}_{(i)}, Q\},
\widehat{R}\}
\end{equation} \vspace{0 mm} \noindent

In fact:

\begin{eqnarray}
&&\{-\widetilde{\cal H}_{\scriptscriptstyle X_{(i)}},\widehat{R}\}=\{-\widetilde{\cal H}_{
\scriptscriptstyle X_{(i)}},\widehat{Y}_{(1)}
\widehat{Y}_{(2)}\cdots
\widehat{Y}_{(r)}\}=\{-\widetilde{\cal H}_{\scriptscriptstyle X_{(i)}}, \widehat{Y}_{(1)}\}
\widehat{Y}_{(2)}\cdots
\widehat{Y}_{(r)}
+\nonumber\\
&&+\widehat{Y}_{(1)}\{-\widetilde{\cal H}_{\scriptscriptstyle X_{(i)}}, 
\widehat{Y}_{(2)}\}\widehat{Y}_{(3)}\cdots
\widehat{Y}_{(r)}+\cdots+\widehat{Y}_{(1)}\widehat{Y}_{(2)}\cdots\widehat{Y}_{(r-1)}
\{-\widetilde{\cal H}_{\scriptscriptstyle X_{(i)}}, \widehat{Y}_{(r)}\}=\nonumber\\
&&=\sum_{j=1}^r\widehat{Y}_{(1)}\widehat{Y}_{(2)}\cdots([X_{(i)},Y_{(j)}])^{\wedge}\cdots
\widehat{Y}_{(r)}=({\cal L}_{\scriptscriptstyle X_{(i)}}R)^{\wedge} 
\end{eqnarray} \vspace{0 mm} \noindent

We note that the extended Poisson brackets take automatically into account 
the sum over $j$ which appears 
in the definition of Lie derivative of a multivector.

Now we can cosider the SN brackets. According to their definition we have

\begin{equation}
[P,R]_{\scriptscriptstyle SN} \; \hat{\longrightarrow} \; 
\sum_{i=1}^p(-1)^{i+1}\widehat{X}_{(1)}\cdots{\widehat {\widehat  X}}_{(i)}
\cdots\widehat{X}_{(p)}\{-\{\widehat{X}_{(i)}, Q\},\widehat{R}\} \label{SNlong}
\end{equation} \vspace{0 mm} \noindent

The previous formula can be written in a very compact way as:

\begin{equation}
\label{compact}
[P,R]_{\scriptscriptstyle SN} \;\; \hat{\longrightarrow} \;\; -\{\{Q,\widehat {P}\},\widehat {R}\}
\end{equation} \vspace{0 mm} \noindent 

In fact:

\begin{eqnarray}
&&-\{\{Q,\widehat {P}\},\widehat {R}\}=-\{\{Q,\widehat{X}_{(1)}\cdots\widehat{X}_{(p)}\},\widehat{R}\}=
\nonumber\\
&&=-\{\{Q,\widehat{X}_{(1)}\}\widehat{X}_{(2)}\cdots\widehat{X}_{(p)},\widehat{R}\}
+\{\widehat{X}_{(1)}\{Q,\widehat{X}_{(2)}\}\widehat{X}_{(3)}\cdots\widehat{X}_{(p)},
\widehat{R}\}-\cdots=\nonumber\\
&&=-\{{\widehat {\widehat  X}}_{(1)}\widehat{X}_{(2)}\cdots\widehat{X}_{(p)}
\{Q,\widehat{X}_{(1)}\},\widehat{R}\}
+\{\widehat{X}_{(1)}{\widehat {\widehat  X}}_{(2)}\cdots\widehat{X}_{(p)}\{Q,\widehat{X}_{(2)}\},
\widehat{R}\}-\cdots=
\nonumber\\
&&=\sum_{i=1}^p(-)^{i+1}\widehat{X}_{(1)}\cdots{\widehat {\widehat  X}}_{(i)}
\cdots\widehat{X}_{(p)}\{-\{Q,\widehat{X}_{(i)}\},\widehat{R}\}=[P,R]^{\wedge}_
{\scriptscriptstyle SN}
\end{eqnarray} \vspace{0 mm} \noindent

So we don't need any sum or any strange factor if we use the {\it EPB} brackets 
to represent the SN brackets. On the RHS of (\ref{compact}) we have the images, via the $\wedge$-map,
of the multivectors $P$ and $R$, which appear on the LHS of the same equation, 
and the usual BRS charge $Q$ that appears naturally also in this context. 

Like in the case of vector fields, where 
$\widetilde{\cal H}_{\scriptscriptstyle X}=\{\widehat{X},Q\}=
\{Q,\widehat{X}\}$,
we can also define a sort of Hamiltonian associated with a multivector 
in the following way:

\begin{equation}
\widetilde{\cal H}_{\scriptscriptstyle P}=\{Q,\widehat{X}_{(1)}\cdots\widehat{X}_{(p)}\}
=\sum_{i=1}^p(-1)^{i+1}\widehat{X}_{(1)}\cdots{\widehat{\widehat X}}_{(i)}\cdots\widehat{X}_{(p)}
\widetilde{\cal H}_{\scriptscriptstyle X_{(i)}} \label{generelazing1}
\end{equation} \vspace{0 mm} \noindent

In this way we can finally write:

\begin{equation}
[P,R]_{\scriptscriptstyle SN} \;\; \hat{\longrightarrow} \;\; -\{\widetilde{\cal H}_
{\scriptscriptstyle P},\widehat {R}\} \label{generelazing2}
\end{equation} \vspace{0 mm} \noindent

From the expressions (\ref{generelazing1}) and (\ref{generelazing2}) we also 
notice  how 
the SN brackets become the usual Lie brackets in case of vector fields. 

Besides this, we can use the properties
of the {\it EPB}  and of Grassmannian variables to demonstrate immediately 
some other properties, or 
alternative definitions, of the Schouten-Nijenhuis brackets. 
If we start from (\ref{SNlong}) and we take into account the definition of Lie derivative
of a multivector, we obtain:

\begin{eqnarray}
&&[P,R]_{\scriptscriptstyle SN} \; \hat{\longrightarrow} \; \sum_{i=1}^p(-1)^{i+1}\widehat{X}_{(1)}
\cdots{\widehat {\widehat  X}}_{(i)}\cdots\widehat{X}_{(p)}\{-\widetilde{\cal H}_{\scriptscriptstyle X_{(i)}},
\widehat{R}\}\nonumber\\
&&=\sum_{i=1}^p\sum_{j=1}^r(-1)^{i+1}\widehat{X}_{(1)}\cdots{\widehat {\widehat  X}}_{(i)}\cdots
\widehat{X}_{(p)}\widehat{Y}_{(1)}\cdots\widehat{Y}_{(j-1)}
([X_{(i)},Y_{(j)}])^{\wedge}\cdots\widehat{Y}_{(r)}
\end{eqnarray} \vspace{0 mm} \noindent

Remembering that the Lie bracket of two vector fields is a vector field 
(and so it is  Grassmannian odd in our language) we can write:

\begin{eqnarray}
\displaystyle
&&[P,R]_{\scriptscriptstyle SN} \; \hat{\longrightarrow} \; \sum_{i=1}^p\sum_{j=1}^r(-1)^{i+j+p+1}
([X_{(i)},Y_{(j)}])^{\wedge}\widehat{X}_{(1)}\cdots{\widehat {\widehat  X}}_{(i)}\cdots
\widehat{X}_{(p)}\widehat{Y}_{(1)}\cdots{\widehat {\widehat  Y}}_{(j)}\cdots\widehat{Y}_{(r)}\nonumber\\
&&=\Biggl((-1)^{p+1}\sum_{i=1}^p\sum_{j=1}^r(-1)^{i+j}[X_{(i)},Y_{(j)}]\wedge X_{(1)}\wedge
\cdots\wedge{\widehat {\widehat  X}}_{(i)}\wedge\cdots\wedge X_{(p)}\wedge Y_{(1)}\wedge\nonumber\\
&&\;\;\;\;\;\qquad\qquad \wedge\cdots\wedge
{\widehat {\widehat  Y}}_{(j)}\wedge\cdots\wedge Y_{(r)}\Biggr)^{\wedge}
\end{eqnarray} \vspace{0 mm} \noindent

In the same way we can start from the last equation and we can use properties of Grassmannian 
variables to obtain the formula:

\begin{eqnarray}
\displaystyle
&&\qquad\qquad\qquad [P,R]_{\scriptscriptstyle SN} \; \hat{\longrightarrow} \; \nonumber\\
&&\hat{\longrightarrow} \; (-1)^{pr}
\sum_{i=1}^p\sum_{j=1}^r(-1)^{i+j}\widehat{Y}_{(1)}\cdots
{\widehat {\widehat  Y}}_{(j)}\cdots\widehat{Y}_{(r)}[Y_{(j)},X_{(i)}]^{\wedge}
\widehat{X}_{(1)}\cdots{\widehat {\widehat  X}}_{(i)}\cdots\widehat{X}_{(p)}=\nonumber\\
&&=(-1)^{pr}\sum_{j=1}^r(-1)^{i+j}\widehat{Y}_{(1)}\cdots
{\widehat {\widehat  Y}}_{(j)}\cdots\widehat{Y}_{(r)}\sum_{i=1}^p(-1)^{i-1}\widehat{X}_{(1)}\cdots
([Y_{(j)},X_{(i)}])^{\wedge}\cdots\widehat{X}_{(p)}=\nonumber\\
&&=\Biggl((-1)^{pr}\sum_{j=1}^r(-1)^{j+1}Y_{(1)}\wedge\cdots\wedge
{\widehat {\widehat  Y}}_{(j)}\wedge\cdots\wedge Y_{(r)}({\cal L}_{\scriptscriptstyle Y_j}P)
\Biggr)^{\wedge}
\end{eqnarray} \vspace{0 mm} \noindent

In this way we have obtained two other properties of the SN brackets that may 
be considered as alternative definitions of the brackets themselves, 
as one can see from \cite{vaisman}. 

\newpage
\section*{APPENDIX C}
{\centerline{\bf CALCULATIONAL DETAILS REGARDING THE FN BRACKETS}

In this section we will handle vector-valued forms $K\in\Omega^{k+1}(M;TM)$. Usually we 
indicate $(k+1)$-forms with $\Omega^{k+1}(M)$, but when we indicate in $\Omega$ also $TM$ we mean
vector-valued forms. Via our $\wedge$-map $K$ becomes:

\begin{equation}
\displaystyle
K \; \hat{\longrightarrow} \; \frac{1}{(k+1)!}\;K^{i}_{i_1i_2\cdots i_{k+1}}
\;[c^{i_1}c^{i_2}\cdots c^{i_{k+1}}][\bar{c}_i] \label{Kmap}
\end{equation} \vspace{0 mm} \noindent

Following \cite{Kolar} we can introduce the interior product between vector-valued forms and usual
forms $\omega$. If $K\in\Omega^{k+1}(M;TM)$ and $\omega\in\Omega^{l}(M)$, then $\iota_
{\scriptscriptstyle K}\omega$ is a $(k+l)$-form, so it can eat multivectors of degree $k+l$ and
it is defined as:

\begin{eqnarray}
\label{Kcontrdef}
&&\qquad \qquad \qquad (\iota_{\scriptscriptstyle K}\omega)[X_{(1)},\cdots, X_{(k+l)}]\equiv\\
&&{1\over (k+1)!(l-1)!}\sum_{\scriptscriptstyle\{\sigma\in S_{k+l}\}}
sign\,\sigma\,\omega[K(X_{\sigma(1)},
\cdots,X_{\sigma(k+1)}), X_{\sigma
(k+2)},\cdots,X_{\sigma(k+l)}]\nonumber 
\end{eqnarray} \vspace{0 mm} \noindent

Which is the $\wedge$-map of $\iota_{\scriptscriptstyle K}\omega$? We expect, as in the case
of interior contraction with vector fields, that:

\begin{equation}
\iota_{\scriptscriptstyle K}\omega \; \hat{\longrightarrow} \; i\{\widehat{K},\widehat{\omega}\}
\label{Kcontr}
\end{equation} \vspace{0 mm} \noindent

but we have to control that (\ref{Kcontr}) is in accordance with the general 
definition (\ref{Kcontrdef}). If $K=\alpha\otimes X$ then we can rewrite 
(\ref{Kcontrdef}) 
in the following way:

\begin{eqnarray}
\displaystyle
&&\qquad \qquad \qquad (\iota_{\scriptscriptstyle K}\omega)[X_{(1)},\cdots, X_{(k+l)}]\equiv\nonumber\\
&&\equiv\frac{1}{(k+1)!(l-1)!}\sum_{\scriptscriptstyle\{\sigma\in S_{k+l}\}}
sign\,\sigma\,\omega[\alpha(X_{\sigma(1)},
\cdots,X_{\sigma(k+1)})X, X_{\sigma
(k+2)},\cdots,X_{\sigma(k+l)}]=\nonumber\\
&&=\frac{1}{(k+1)!(l-1)!}\sum_{\scriptscriptstyle\{\sigma\in S_{k+l}\}}
sign\,\sigma\,\alpha(X_{\sigma(1)},
\cdots,X_{\sigma(k+1)})\omega[X, X_{\sigma
(k+2)},\cdots,X_{\sigma(k+l)}]=\nonumber\\
&&=\frac{1}{(k+1)!(l-1)!}\sum_{\scriptscriptstyle\{\sigma\in S_{k+l}\}}
sign\,\sigma\,\alpha(X_{\sigma(1)},
\cdots,X_{\sigma(k+1)})\iota_{\scriptscriptstyle X}[X_{\sigma
(k+2)},\cdots,X_{\sigma(k+l)}]=\nonumber\\
&&=\alpha\wedge\iota_{\scriptscriptstyle X}\omega(X_{(1)},\cdots,X_{(k+l)})
\end{eqnarray} \vspace{0 mm} \noindent

where $\alpha\in\Omega^{k+1}(M)$ and $\iota_{\scriptscriptstyle X}\omega\in\Omega^{l-1}(M)$.
From the previous equalities we deduce that we can translate the interior contraction
between a form and a vector-valued form as the exterior product between two 
forms:

\begin{equation}
\iota_{\scriptscriptstyle K}\omega \;\; = \;\; \alpha\wedge\iota_{\scriptscriptstyle X}\omega
\end{equation} \vspace{0 mm} \noindent

Now we want to prove that, if $\alpha\in\Omega^{k+1}(M)$ and $\beta\in\Omega^{l-1}(M)$ 
are two differential 
forms, we can represent their exterior product as:

\begin{equation}
(\alpha\wedge\beta) \;\; \hat{\longrightarrow} \;\; \widehat{\alpha}\widehat{\beta} \label{alphabeta}
\end{equation} \vspace{0 mm} \noindent

The definition of the exterior product of differential forms is:

\begin{eqnarray}
\displaystyle 
\label{exteriorproduct}
&&\qquad\qquad\qquad\alpha\wedge\beta(X_{(1)},\cdots,X_{(k+l)})\equiv\\
&&\equiv\frac{1}{(k+1)!(l-1)!}\sum_{\{\sigma\in S_{k+l}\}} sign\,
\sigma\,\alpha(X_{\sigma(1)},\cdots,X_{\sigma(k+1)})\beta(X_{\sigma(k+2)},\cdots,X_{\sigma(k+l)})
\nonumber
\end{eqnarray} \vspace{0 mm} \noindent

We can start translating via the $\wedge$-map the RHS of 
(\ref{exteriorproduct}):

\begin{eqnarray}
\displaystyle
&&\alpha(X_{\sigma(1)},\cdots,X_{\sigma(k+1)})=\iota_{\scriptscriptstyle X_{\sigma(k+1)}}\cdots
\iota_{\scriptscriptstyle X_{\sigma(1)}}\alpha \; \hat{\longrightarrow}\nonumber\\
&&\hat{\longrightarrow} \;
i^{k+1}\{X^{j_{k+1}}_{\sigma(k+1)}\bar{c}_{j_{k+1}}\{\cdots,\{X^{j_1}_{\sigma(1)}
\bar{c}_{j_1},\widehat{\alpha}\underbrace{\}\}\cdots\}}_{\scriptscriptstyle k+1 \,brackets}=
\\
&&=\frac{1}{(k+1)!}\sum_{\{\tau\in S_{k+1}\}}\,sign\,\tau \,X^{j_{k+1}}_{\sigma(k+1)}\cdots
X^{j_1}_{\sigma(1)}\alpha_{\tau(j_1)\cdots\tau(j_{k+1})}=X^{j_{k+1}}_{\sigma(k+1)}\cdots
X^{j_1}_{\sigma(1)}\alpha_{j_1\cdots j_{k+1}}
\nonumber
\end{eqnarray} \vspace{0 mm} \noindent

In the last passage we have employed the fact that $\alpha_{\tau(j_1)\cdots\tau(j_{k+1})}=
sign\,\tau\,\alpha_{j_1\cdots j_{k+1}}$ and that the number of permutations in $S_{k+1}$
is just $(k+1)!$.
In the same way we have:

\begin{equation}
\beta(X_{\sigma(k+2)},\cdots,X_{\sigma(k+l)}) \; \hat{\longrightarrow} \;
X^{j_{k+l}}_{\sigma(k+l)}\cdots
X^{j_{k+2}}_{\sigma(k+2)}\beta_{j_{k+2}\cdots j_{k+l}}
\end{equation} \vspace{0 mm} \noindent

So, from the definition itself of exterior product, we must have:

\begin{eqnarray}
\displaystyle
\label{alphabeta2}
\alpha\wedge\beta(X_{(1)},\cdots,X_{(k+l)}) \; & \hat{\longrightarrow} & \;
\frac{1}{(k+1)!(l-1)!}\sum_{\{\sigma\in S_{k+l}\}}sign\,\sigma\,X^{j_{k+l}}_{\sigma(k+l)}\cdots
X^{j_{k+2}}_{\sigma(k+2)}\cdot\nonumber\\
&&X^{j_{k+1}}_{\sigma(k+1)}\cdots
X^{j_1}_{\sigma(1)}\alpha_{j_1\cdots j_{k+1}}\beta_{j_{k+2}\cdots j_{k+l}}
\end{eqnarray} \vspace{0 mm} \noindent

Now we can say that (\ref{alphabeta}) is correct if it reproduces (\ref{alphabeta2}).
So we have to evaluate:

\begin{eqnarray}
\displaystyle
\label{alphabeta3}
&&\alpha\wedge\beta(X_{(1)},\cdots,X_{(k+l)}) \; \hat{\longrightarrow} \;
\frac{1}{(k+1)!(l-1)!}i^{k+l}\{X^{i_{k+l}}_{(k+l)}\bar{c}_{i_{k+l}},\{\cdots,\{X^{i_1}_{(1)}\bar{c}
_{i_1},\nonumber\\
&&\alpha_{j_1\cdots j_{k+1}}c^{j_1}\cdots c^{j_{k+1}}\beta_{j_{k+2}\cdots j_{k+l}}
c^{j_{k+2}}\cdots c^{j_{k+l}}\underbrace{\}\}\cdots\}}_{\scriptscriptstyle k+l\,brackets}
=\nonumber\\
&&=\frac{1}{(k+1)!(l-1)!}\sum_{\{\sigma\in S_{k+l}\}}sign\,\sigma\,
X^{\sigma(j_{k+l})}_{(k+l)}\cdots X^{\sigma(j_{1})}_{(1)}\alpha_{j_1\cdots j_{k+1}}
\beta_{j_{k+2}\cdots j_{k+l}}=\nonumber\\
&&=\frac{1}{(k+1)!(l-1)!}\sum_{\{\sigma\in S_{k+l}\}}sign\,\sigma\,
X^{j_{k+l}}_{\sigma(k+l)}\cdots X^{j_{1}}_{\sigma(1)}\alpha_{j_1\cdots j_{k+1}}
\beta_{j_{k+2}\cdots j_{k+l}}  
\end{eqnarray} \vspace{0 mm} \noindent

The last terms of (\ref{alphabeta2}) and (\ref{alphabeta3}) are equal; 
so we can conclude
that the correct rappresentation via $\wedge$-map of $\alpha\wedge\beta$ 
is just
$\widehat{\alpha}\widehat{\beta}$. 

At this point we have all the elements to translate in our language the operation $\iota_{
\scriptscriptstyle K}\omega$.
In fact:

\begin{equation}
\iota_{\scriptscriptstyle K}\omega \;\; = \;\; \alpha\wedge\iota_{\scriptscriptstyle X}\omega \;\; 
\hat{\longrightarrow} \;\; \widehat{\alpha}
(\iota_{\scriptscriptstyle X}\omega)^{\wedge}
\end{equation} \vspace{0 mm} \noindent

Using (\ref{eq:A-nove}) and the fact that $\{\widehat{\alpha},\widehat{\omega}\}=0$ we can go on 
writing:

\begin{equation}
\widehat{\alpha}(\iota_{\scriptscriptstyle 
X}\omega)^{\wedge}=i\widehat{\alpha}\{\widehat{X},\widehat{\omega}\}=
i\{\widehat{\alpha}\widehat{X},\widehat{\omega}\}=i\{\widehat{K},\widehat{\omega}\}
\end{equation} \vspace{0 mm} \noindent

So, as we had expected, we have proved that:

\begin{equation}
\iota_{\scriptscriptstyle K}\omega \;\;\; \hat{\longrightarrow} \;\;\; 
i\{\widehat{K},\widehat{\omega}\}
\end{equation} \vspace{0 mm} \noindent

At this point, having defined the concept of interior contraction with a
vector-valued form, we can go on introducing the Lie derivative associated with a vector valued 
form $K$:

\begin{equation}
{\cal L}_{\scriptscriptstyle K}\;\;=\;\;[\iota_{\scriptscriptstyle K},d]
\;\;=\;\;\iota_{\scriptscriptstyle K}d
+(-1)^{k+1}d\iota_{\scriptscriptstyle K}
\end{equation} \vspace{0 mm} \noindent

Since we know how to translate  in our language both the interior contraction and the exterior
derivative, we can write:

\begin{eqnarray}
&&{\cal L}_{\scriptscriptstyle K}\omega \;\; \hat{\longrightarrow} \;\; 
i\{\widehat{K},(d\omega)^{\wedge}
\}+(-1)^{k+1}i\{Q,(\iota_{\scriptscriptstyle K}\omega)^{\wedge}\}=\nonumber\\
&&=-\{\widehat{K},\{Q,\widehat{\omega}\}\}+(-1)^{k}\{Q,\{\widehat{K},\widehat{\omega}\}\}
=-\{\{\widehat{K},Q\},\widehat{\omega}\}
\end{eqnarray} \vspace{0 mm} \noindent

where, in the last step, we have used the Jacobi identity.
So we have:

\begin{equation}
{\cal L}_{\scriptscriptstyle K}\omega \; \hat{\longrightarrow} \; -\{\widetilde{\cal H}_
{\scriptscriptstyle K},\widetilde{\omega}\} \label{ellacca}
\end{equation} \vspace{0 mm} \noindent

where we have  defined, as usual, $\widetilde{\cal H}_{\scriptscriptstyle K}=\{\widehat{K},Q\}$.
From the definition itself and making use of (\ref{Kmap}) it follows that the explicit expression of 
$\widetilde{\cal H}_
{\scriptscriptstyle K}$ is:

\begin{equation}
\displaystyle
\widetilde{\cal H}_{\scriptscriptstyle K}=\frac{1}{(k+1)!}\Biggl(\lambda_jK^j_{j_1j_2\cdots j_{k+1}}
+i\bar{c}_j(\partial_dK^{j}_{j_1j_2\cdots j_{k+1}}c^d)\Biggr)c^{j_1}\cdots c^{j_{k+1}}
\end{equation} \vspace{0 mm} \noindent

From the previous expression we note that if $k$ is even then $\widetilde{\cal H}_
{\scriptscriptstyle K}$ is Grassmannian odd and if $k$ is odd $\widetilde{\cal H}_
{\scriptscriptstyle K}$ is even. Moreover, from (\ref{ellacca}), the Grassmannian parity 
of $\widetilde{\cal H}_{\scriptscriptstyle K}$ coincides 
with that of the correspondent Lie derivative ${\cal L}_{\scriptscriptstyle K}$.

Finally we have all the elements to translate in our language the Fr\"olicher-Nijenhuis brackets.
They are defined~\cite{Kolar} in implicit way from the equation:

\begin{equation}
[{\cal L}_{\scriptscriptstyle K},{\cal L}_{\scriptscriptstyle L}] \;\; = \;\;
{\cal L}_{\scriptscriptstyle
[K,L]_{FN}} \label{FNdef}
\end{equation} \vspace{0 mm} \noindent

Now if we think of LHS of (\ref{FNdef}) as applied on a generic 
form $\omega$ we have:

\begin{equation}
[{\cal L}_{\scriptscriptstyle K},{\cal L}_{\scriptscriptstyle L}]\omega=
({\cal L}_{\scriptscriptstyle K}{\cal L}_{\scriptscriptstyle L})\omega
-(-1)^{[\widetilde{\cal H}_{\scriptscriptstyle L}][\widetilde{\cal H}_{\scriptscriptstyle K}]}
({\cal L}_{\scriptscriptstyle L}{\cal L}_{\scriptscriptstyle K})\omega
\end{equation} \vspace{0 mm} \noindent

where we indicate with $[(\;\cdot\;)]$ the Grassmannian parity of $(\;\cdot\;)$.
Via our mapping we have:

\begin{eqnarray}
\label{comparison1}
&&[{\cal L}_{\scriptscriptstyle K},{\cal L}_{\scriptscriptstyle L}]\omega \; \hat{\longrightarrow} \;
\{\widetilde{\cal H}_{\scriptscriptstyle K},\{\widetilde{\cal H}_{\scriptscriptstyle L},
\widehat{\omega}\}\}-(-1)^{[\widetilde{\cal H}_{\scriptscriptstyle L}][\widetilde{\cal H}_
{\scriptscriptstyle K}]}\{\widetilde{\cal H}_{\scriptscriptstyle L},
\{\widetilde{\cal H}_{\scriptscriptstyle K},\widehat{\omega}\}\}=\nonumber\\
&&=\{\widetilde{\cal H}_{\scriptscriptstyle K},\{\widetilde{\cal H}_{\scriptscriptstyle L},
\widehat{\omega}\}\}+(-1)^{[\widetilde{\cal H}_{\scriptscriptstyle K}]([\widetilde{\cal H}_
{\scriptscriptstyle L}]+[\omega])}\{\widetilde{\cal H}_{\scriptscriptstyle L},
\{\omega, \widetilde{\cal H}_{\scriptscriptstyle K}\}\}=\{\{\widetilde{\cal H}_
{\scriptscriptstyle K},\widetilde{\cal H}_{\scriptscriptstyle L}\},\widehat{\omega}\}
\end{eqnarray} \vspace{0 mm} \noindent

where in the last step we have used, as usual, the Jacobi identity.
The RHS of (\ref{FNdef}) can be translated as:

\begin{equation}
\label{comparison2}
{\cal L}_{\scriptscriptstyle [K,L]_{FN}}\omega \; \hat{\longrightarrow} \;
-\{\widetilde{\cal H}_{\scriptscriptstyle [K,L]_{FN}},\hat{\omega}\}
\end{equation} \vspace{0 mm} \noindent

so, from the comparison of (\ref{comparison1}) and (\ref{comparison2}),
we have the following important relation:

\begin{equation}
\label{FNimportant}
\widetilde{\cal H}_{\scriptscriptstyle [K,L]_{FN}}=\{([K,L]_{\scriptscriptstyle FN})^{\wedge},Q\}
=-\{\widetilde{\cal H}_
{\scriptscriptstyle K},\widetilde{\cal H}_{\scriptscriptstyle L}\}
\end{equation} \vspace{0 mm} \noindent

Now, if we want to have the correct representation of the FN brackets, we have to write
$\{\widetilde{\cal H}_
{\scriptscriptstyle K},\widetilde{\cal H}_{\scriptscriptstyle L}\}$ as 
$\{(\cdot), Q\}$.
This is not difficult to do, in fact:

\begin{equation}
\label{Qform}
\{\widetilde{\cal H}_
{\scriptscriptstyle K},\widetilde{\cal H}_{\scriptscriptstyle L}\}
\;\; = \;\; \{\{\{\widehat{K},Q\},\widehat{L}\},Q\} 
\end{equation} \vspace{0 mm} \noindent

To demonstrate (\ref{Qform}) we can start from its RHS and employ the 
Jacobi identity:

\begin{equation}
\{\{\{\widehat{K},Q\},\widehat{L}\},Q\} =
\{\{\widetilde{\cal H}_{\scriptscriptstyle K}, \widehat{L}\}, Q\}=\{\widetilde{\cal H}_
{\scriptscriptstyle K},\widetilde{\cal H}_{\scriptscriptstyle L}\}-(-1)^{k+l}
\{\{Q,\widetilde{\cal H}_{\scriptscriptstyle K}\},L\}
\end{equation} \vspace{0 mm} \noindent

So (\ref{Qform}) is proved if $\{Q,\widetilde{\cal H}_{\scriptscriptstyle K}\}=0$.
But this is easy to demonstrate, since every BRS exact term has zero {\it EPB}
with Q. In fact:

\begin{equation}
\{Q,\widetilde{\cal H}_{\scriptscriptstyle K}\}=\{Q,\{\widehat{K},Q\}\}
\end{equation} \vspace{0 mm} \noindent

The Jacobi identity in this case is:

\begin{equation}
\{Q,\{\widehat{K},Q\}\}+\{Q,\{\widehat{K},Q\}\}+\{\widehat{K},\{Q,Q\}\}=0
\end{equation} \vspace{0 mm} \noindent

From the nilpotency of $Q$ we can conclude that:

\begin{equation}
\{Q,\{\widehat{K},Q\}\}=\{Q,\widetilde{\cal H}_{\scriptscriptstyle K}\}=0
\end{equation} \vspace{0 mm} \noindent

and so (\ref{Qform}) is proved. Substituting (\ref{Qform}) into 
(\ref{FNimportant}) we obtain finally:

\begin{equation}
[K,L]_{\scriptscriptstyle FN} \; \hat{\longrightarrow} \; 
=-\{\widetilde{\cal H}_{\scriptscriptstyle K},\widehat{L}\}
\end{equation} \vspace{0 mm} \noindent

From the previous expression we notice  how, if $K$ and $L$ are zero vector-valued forms, i.e.
if they are vector fields, then the FN brackets reduce to the usual Lie brackets. In a certain sense
we can say that, as the  SN brackets generalize Lie brackets in the case of 
multivector fields, so 
the FN brackets generalize the Lie brackets in the case of vector-valued forms. 
 
\newpage
\section*{APPENDIX D}
{\centerline{\bf CALCULATIONAL DETAILS REGARDING THE NR BRACKETS}}

The Nijenhuis-Richardson brackets are defined between two vector-valued forms:\break 
$K\in\Omega^{k+1}
(M;TM)$ and $L\in\Omega^{l+1}(M;TM)$ and they give a vector-valued form of 
degree $k+l+1$ defined~\cite{Kolar} 
in an implicit way as:

\begin{equation}
\iota_{\scriptscriptstyle [K,L]_{NR}} \;\;\; \equiv \;\;\; [\iota_{\scriptscriptstyle K},
\iota_{\scriptscriptstyle L}] \label{NRdef}
\end{equation} \vspace{0 mm} \noindent

If we apply  a generic form $\omega\in\Omega^m(M)$ on the  LHS of (\ref{NRdef})
then, via our $\wedge$-map, it becomes:
 
\begin{equation}
\iota_{\scriptscriptstyle [K,L]_{NR}}\omega \;\;\; \hat{\longrightarrow} \;\;\; i\{([K,L]_
{\scriptscriptstyle NR})^{\wedge},\widehat{\omega}\} \label{alpha}
\end{equation} \vspace{0 mm} \noindent

while the RHS of (\ref{NRdef}) becomes:

\begin{eqnarray}
&&[\iota_{\scriptscriptstyle K},\iota_{\scriptscriptstyle L}]\omega=
\iota_{\scriptscriptstyle K}(\iota_{\scriptscriptstyle L}\omega)-(-1)^{kl}
\iota_{\scriptscriptstyle L}(\iota_{\scriptscriptstyle K}\omega) \;
\hat{\longrightarrow} \; i\{\widehat{K},(\iota_{\scriptscriptstyle L}\omega)^{\wedge}\}
-(-1)^{kl}i\{\widehat{L},(\iota_{\scriptscriptstyle K}\omega)^{\wedge}\}=\nonumber\\
&&-\{\widehat{K},\{\widehat{L},\widehat{\omega}\}\}+(-1)^{kl}\{\widehat{L},\{\widehat{K},
\widehat{\omega}\}\}
\end{eqnarray} \vspace{0 mm} \noindent

Using the Jacobi identity we obtain:

\begin{equation}
[\iota_{\scriptscriptstyle K},\iota_{\scriptscriptstyle L}]\omega \;\;\; \hat{\longrightarrow} \;\;\;
-\{\{\widehat{K},\widehat{L}\},\widehat{\omega}\} \label{beta}
\end{equation} \vspace{0 mm} \noindent

We can write, from the comparison of (\ref{alpha}) with (\ref{beta}):

\begin{equation}
[K,L]_{\scriptscriptstyle NR} \;\;\; \hat{\longrightarrow} \;\;\; i\{\widehat{K},\widehat{L}\}
\label{gamma}
\end{equation} \vspace{0 mm} \noindent

So the NR brackets between two vector-valued forms are just proportional to the extended Poisson
brackets of the vector-valued forms themselves.

Now we can use properties of the extended Poisson brackets to find a more explicit definition of NR
brackets, in fact:

\begin{eqnarray}
\displaystyle
&&i\{\widehat{K},\widehat{L}\}=\frac{i}{(l+1)!(k+1)!}\{K^{i}_{i_1\cdots i_{k+1}}c^{i_1}\cdots
c^{i_{k+1}}\bar{c}_i,L^j_{j_1\cdots j_{l+1}}c^{j_1}\cdots c^{j_{l+1}}\bar{c}_j\}=\nonumber\\
&&=\frac{i}{(l+1)!}\{\widehat{K},L^j_{j_1\cdots j_{l+1}}c^{j_1}\cdots c^{j_{l+1}}\}\bar{c}_j
+\frac{i}{(l+1)!(k+1)!}L^j_{j_1\cdots j_{l+1}}c^{j_1}\cdots c^{j_{l+1}}\nonumber\\
&&\{K^{i}_{i_1\cdots i_{k+1}}
c^{i_1}\cdots c^{i_{k+1}}\bar{c}_i,\bar{c}_j\}(-1)^{(l+1)k}=\frac{i}{(l+1)!}\{\widehat{K},
L^j_{j_1\cdots j_{l+1}}c^{j_1}\cdots c^{j_{l+1}}\}\bar{c}_j+\nonumber\\
&&-\frac{i}{(l+1)!(k+1)!}L^j_{j_1\cdots j_{l+1}}c^{j_1}\cdots c^{j_{l+1}}
\{\bar{c}_j,K^{i}_{i_1\cdots i_{k+1}}c^{i_1}\cdots c^{i_{k+1}}\}\bar{c}_i(-1)^{lk}
=\nonumber\\
&&=\frac{i}{(l+1)!}\{\widehat{K},L^j_{j_1\cdots j_{l+1}}
c^{j_1}\cdots c^{j_{l+1}}\}\bar{c}_j-(-1)^{lk}\frac{i}{(k+1)!}\{\widehat{L},K^{i}_{i_1\cdots i_{k+1}}
c^{i_1}\cdots c^{i_{k+1}}\}\bar{c}_i\nonumber\\
&&
\end{eqnarray} \vspace{0 mm} \noindent

Since $\iota_{\scriptscriptstyle K}(\omega\otimes X)\equiv\iota_{\scriptscriptstyle K}(\omega)
\otimes X$ we can write:

\begin{equation}
i\{\widehat{K},\widehat{L}\}=(\iota_{\scriptscriptstyle K}L)^{\wedge}
-(-1)^{lk}(\iota_{\scriptscriptstyle L}K)^{\wedge} \label{proprNR}
\end{equation} \vspace{0 mm} \noindent

From the comparison of (\ref{gamma}) with (\ref{proprNR}) we obtain:

\begin{equation}
[K,L]_{\scriptscriptstyle NR}=\iota_{\scriptscriptstyle K}L-(-1)^{kl}
\iota_{\scriptscriptstyle L}K
\end{equation} \vspace{0 mm} \noindent

that can be interpreted as a more explicit definition of NR brackets (see also \cite{Kolar}).
\newpage

\section*{Acknowledgements}

The project contained here has been suggested by G. Marmo in June 1994.
We wish to warmly thank him for this and for his patience in waiting
for the tex file of this work. We also wish to thank M. Regini and M. Reuter  
for  several technical suggestions. This work has been
supported by MURST (Italy) and NATO grants.


\end{document}